\documentclass[10pt,conference]{IEEEtran}
\IEEEoverridecommandlockouts

\usepackage{cite}
\usepackage{bm}
\usepackage{amsmath}
\usepackage{extarrows}
\usepackage{amssymb}
\usepackage{graphicx}
\usepackage{color}
\usepackage{enumerate}
\usepackage{bookmark} 
\graphicspath{{figure}}
\usepackage{setspace}
\usepackage{subfigure}
\usepackage{algorithm}
\usepackage{algorithmicx}
\usepackage{multirow}
\usepackage{stfloats}
\usepackage{nicematrix}
\usepackage{pifont}
\usepackage{threeparttable}
\usepackage{verbatim}

\newcommand{\mr}{\mathrm} 

\newcommand{\BE}{\begin{equation}}
\newcommand{\EE}{\end{equation}}
\newcommand{\BS}{\begin{subequations}}
\newcommand{\ES}{\end{subequations}}
\renewcommand{\bf}{\bm}

\newtheorem{theorem}{Theorem}

\newtheorem{lemma}{Lemma}

\allowdisplaybreaks \allowdisplaybreaks[2]

\def\BibTeX{{\rm B\kern-.05em{\sc i\kern-.025em b}\kern-.08em
    T\kern-.1667em\lower.7ex\hbox{E}\kern-.125emX}}

\begin{document}

\title{{Low-Complexity and Information-Theoretic Optimal Memory AMP for Coded Generalized MIMO}}

\author{
\IEEEauthorblockN{
Yufei Chen\IEEEauthorrefmark{1}, \emph{Student Member, IEEE},
Lei Liu\IEEEauthorrefmark{2}, \emph{Senior Member, IEEE}, 
Yuhao Chi\IEEEauthorrefmark{1}, \emph{Member, IEEE}, \\ 
Ying Li\IEEEauthorrefmark{1}, \emph{Member, IEEE}, and 
Zhaoyang Zhang\IEEEauthorrefmark{2}, \emph{Senior Member, IEEE}}

\IEEEauthorblockA{\IEEEauthorrefmark{1}State Key Laboratory of ISN, Xidian University, China}
\IEEEauthorblockA{\IEEEauthorrefmark{2}Zhejiang Provincial Key Laboratory of IPCAN, Zhejiang University, China}

\thanks{This work was supported in part by the China National Key R\&D Program under Grant 2021YFA1000500, in part by the National Natural Science Foundation of China under Grants 62131016, 62201424, 61971333, in part by the Natural Science Foundation of Shaanxi Province 2022JQ-703, in part by the Key Research and Development Program of Shaanxi 2023-YBGY-218, and in part by the Fundamental Research Funds for the Central Universities XJS220117. (Yuhao Chi and Ying Li are both the corresponding authors.)} 
}

\maketitle

\begin{abstract}
    This paper considers a generalized multiple-input multiple-output (GMIMO) with practical assumptions, such as massive antennas, practical channel coding, arbitrary input distributions, and general right-unitarily-invariant channel matrices (covering Rayleigh fading, certain ill-conditioned and correlated channel matrices). 
    Orthogonal/vector approximate message passing (OAMP/VAMP) has been proved to be information-theoretically optimal in GMIMO, but it is limited to high complexity. Meanwhile, low-complexity memory approximate message passing (MAMP) was shown to be Bayes optimal in GMIMO, but channel coding was ignored.
    Therefore, how to design a low-complexity and information-theoretic optimal receiver for GMIMO is still an open issue.
    In this paper, we propose an information-theoretic optimal MAMP receiver for coded GMIMO, whose achievable rate analysis and optimal coding principle are provided to demonstrate its information-theoretic optimality. Specifically, state evolution (SE) for MAMP is intricately multi-dimensional because of the nature of local memory detection. To this end, a fixed-point consistency lemma is proposed to derive the simplified variational SE (VSE) for MAMP, based on which the achievable rate of MAMP is calculated, and the optimal coding principle is derived to maximize the achievable rate. Subsequently, we prove the information-theoretic optimality of MAMP.
    Numerical results show that the finite-length performances of MAMP with optimized LDPC codes are about $1.0\sim2.7$~dB away from the associated constrained capacities. It is worth noting that MAMP can achieve the same performance as OAMP/VAMP with $4\text{\textperthousand}$ of the time consumption for large-scale systems.
\end{abstract}

\section{Introduction}
With the rapid development of wireless communications, 6G networks are expected to provide performance superior to 5G and satisfy emerging services and applications\cite{6G1}. Accordingly, data types in various application scenarios become more diverse, and practical communication scenarios are more complex. However, most conventional multiple-input multiple-output (MIMO) technologies are limited to ideal communication assumptions, i.e., a limited number of antennas, no coding constraint, Gaussian signaling, channel state information (CSI) available at the transceiver, and independent identically distributed (IID) channel matrices, which cannot effectively support the complex 6G scenarios. Therefore, a more practical generalized MIMO (GMIMO)\cite{YuhaoTcom2022} is considered in this paper, including: 1) massive antennas, 2) practical channel coding and decoding, 3) arbitrary input distributions, 4) CSI only available at the receiver, and 5) general right-unitarily-invariant channel matrices, covering Rayleigh fading, certain ill-conditioned and correlated channel matrices. Meanwhile, these generalized assumptions bring new challenges to the design of receivers for GMIMO.

\subsection{Advanced AMP-Type Receivers}
Approximate message passing (AMP)-type algorithms have been widely used in MIMO receivers~\cite{YuhaoTcom2022,MaTWC2019}. AMP is a high-efficient signal recovery algorithm with a low-complexity matched filter (MF) for arbitrary input distributions \cite{BayatiTIT2011}. Remarkably, AMP is proved to be Bayes optimal via a scalar recursion called state evolution (SE)\cite{BayatiTIT2011}. However, AMP is only available for IID channel matrices. For more complex non-IID channel matrices, AMP performs poorly or even diverges\cite{Vila2015ICASSP}. 
To address this issue, orthogonal/vector AMP (OAMP/VAMP) is developed in\cite{MaAcess2017,Rangan2019TIT} for right-unitarily-invariant matrices, employing a linear minimum mean-square error (LMMSE) detector to mitigate linear interference and an orthogonalization to overcome the correlation problem during iteration. The Bayes optimality of OAMP/VAMP is proved via the replica methods in~\cite{Tulino2013TIT}. Due to the high-complexity LMMSE, it is difficult to apply OAMP/VAMP effectively to large-scale systems. To address this challenge, a low-complexity convolutional AMP (CAMP)~\cite{Takeuchi2020CAMP} is proposed, which replaces the Onsager term of AMP with a convolution of all preceding messages. However, CAMP converges slowly and even easily diverges for the channel matrices with high condition numbers. Around the same time, another low-complexity memory AMP (MAMP)~\cite{MAMPTIT} is presented and adopts a long-memory MF (LM-MF) utilizing the information from the previous iterations to replace the LMMSE detector of OAMP/VAMP. Meanwhile, MAMP is proved to be Bayes optimal for right-unitarily-invariant matrices.

However, the above mentioned AMP-type algorithms\cite{BayatiTIT2011,MaAcess2017,Rangan2019TIT,MAMPTIT} only focus on the detection of uncoded systems, ignoring the effects of channel coding and decoding with no guarantee of asymptotically error-free recovery.

\vspace{-0.1cm}
\subsection{Information-Theoretic Optimality of AMP-Type Receivers}
For coded MIMO with IID channel matrices and arbitrary input signaling, the achievable rate analysis and information-theoretic (i.e., constrained-capacity) optimality of AMP are presented in \cite{LeiTIT2021} based on the scalar SE. Specifically, the optimal coding principle is derived by tracking the scalar estimated variance between linear detector (LD) and nonlinear detector (NLD) while satisfying the error-free decoding condition. On this basis, in \cite{LeiTIT2021}, the maximum achievable rate of AMP is proved to equal the associated constrained capacity~\cite{Barbier2018b}. For right-unitarily-invariant channel matrices and arbitrary input signaling, OAMP/VAMP is shown to achieve the constrained capacity of GMIMO in \cite{YuhaoTcom2022,LeiOptOAMP}. Unlike AMP, the orthogonalization in the LD and NLD of OAMP/VAMP destroys the MMSE property, making it impossible to directly calculate the achievable rates of OAMP/VAMP based on mutual information-MMSE (I-MMSE) lemma\cite{GuoTIT2005}. To overcome this difficulty, a variational SE (VSE) of OAMP/VAMP is developed by incorporating all orthogonal operations into the LD\cite{LeiOptOAMP}, based on which the achievable rate analysis and optimal coding principle are obtained using I-MMSE lemma.

The information-theoretic optimality of AMP and OAMP/VAMP is restricted to IID channel matrices and high complexity, respectively, which are difficult to apply effectively to large-scale GMIMO. How to design a low-complexity and information-theoretic optimal receiver for GMIMO is still an open issue.

\begin{figure*}\vspace{-0.3cm}
	\centering
	\includegraphics[width=0.8\linewidth]{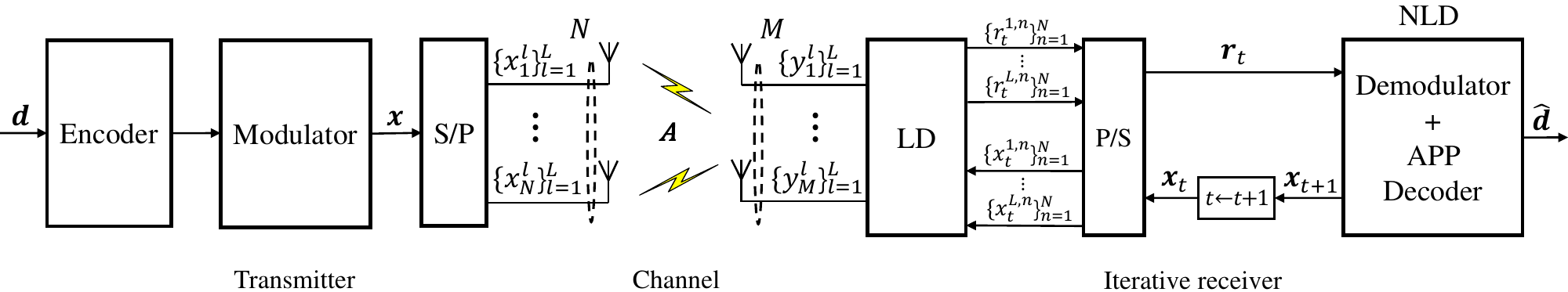}\vspace{-0.15cm}
	\caption{An uplink GMIMO system: an $N$-antennas transmitter and an $M$-antennas iterative receiver consisting of an LD and an NLD. S/P and P/S denote serial-to-parallel and parallel-to-serial conversion, respectively.}\label{Fig:System model}  \vspace{-0.2cm}
\end{figure*}

\vspace{-0.1cm}
\subsection{Motivation and Contributions}
 A promising low-complexity candidate of GMIMO receivers is MAMP, which consists of a memory LD (MLD) and an NLD. However, in\cite{MAMPTIT}, it only focused on the design of Bayes optimal MAMP in uncoded systems while ignoring the effect of channel coding and decoding. Meanwhile, the Bayes optimal MAMP with a well-designed P2P channel code is still suboptimal, as demonstrated in numerical results in this paper. On the other hand, due to the memory involved in local detectors, a covariance-matrix SE containing complex multi-dimensional transfer functions is required for MAMP to evaluate the asymptotic performance. However, the existing achievable rate analysis and optimal coding principle are based on single-input single-output (SISO) transfer functions~\cite{LeiTIT2021,LeiOptOAMP,YuhaoTcom2022}, which are infeasible to be directly extended to~MAMP.
 
To address the above challenges, we propose an MAMP receiver for coded GMIMO, in which the achievable rate analysis and optimal coding principle are provided for MAMP to prove its constrained-capacity optimality.
To avoid the complex multi-dimensional SE analysis of MAMP, a lemma about the SE fixed-point consistency of MAMP and OAMP/VAMP is proposed.
According to this lemma, a simplified SISO VSE for MAMP is derived, based on which the achievable rate of MAMP is obtained.
The optimal coding principle of MAMP is derived with the goal of maximizing the achievable rate, and the maximum achievable rate equals the constrained capacity of GMIMO. Therefore, the constrained-capacity optimality of MAMP is established.
Moreover, we compare the maximum achievable rates of MAMP and the existing cascading MAMP (CAS-MAMP, i.e., with separate MLD and NLD) in GMIMO.
Furthermore, a kind of practical LDPC code is designed for MAMP in GMIMO.
The main contributions of this paper are summarized as follows.
\begin{enumerate}
	\item  A simplified SISO VSE of MAMP is derived, from which the achievable rate analysis and optimal coding principle of MAMP are obtained. 
	\item The constrained-capacity optimality of MAMP is proved, i.e., the maximum achievable rate of MAMP is equal to the constrained capacity of GMIMO.
    \item Taking the ill-conditioned channel matrices as examples, the maximum achievable rates of MAMP are presented, along with a comparison to the existing CAS-MAMP. 
	\item A kind of capacity-approaching LDPC code is designed for MAMP. Numerical results show that the finite-length performances of MAMP with optimized LDPC codes are about $1.0\sim2.7$~dB away from the associated constrained capacities. It is worth noting that MAMP only takes $4\text{\textperthousand}$ of the execution time of OAMP/VAMP to achieve the same performance for large-scale systems.
\end{enumerate}

\emph{Note:} Due to the limitation of pages, detailed proofs of Lemmas \ref{lem:same_fp} and \ref{lem:VSE_MAMP} are given in a full version of this paper.

\section{System Model}
Fig.~\ref{Fig:System model} illustrates an uplink GMIMO system with an $N$-antennas transmitter and one receiver equipped with $M$ antennas. In the transmitter, message sequence $\bf{d}$ is encoded by a forward error control (FEC) encoder. After modulation, length-$NL$ modulated sequence $\bf{x}$ is generated and transformed into $N$ sequences $\{x^l_{n}\}_{l=1}^L, n=1,...,N,$ by serial-to-parallel conversion, in which each entry of $\bf{x}$ is taken from a discrete constellation set $\cal{S}$. At the $l$-th time slot, symbol sequence $\bf{x}^l=[x^l_{1},...,x^l_{N}]^{\rm{T}}$ is transmitted to the channel,  satisfying the power constraint $\tfrac{1}{N}\mr{E}\{\|\bf{x}^l\|^2\}=1$.

The receiver obtains signal $\bf{y}^l= [y_1^l,...,y_M^l]^{\rm{T}}$ given by
\BE\label{Eqn:y^t}
\bf{y}^l=\bf{A}\bf{x}^l+\bf{n}^l,\;\;  l=1,\dots,L,
\EE
where $\bf{A}\in \mathbb{C}^{M\times N}$ is a channel matrix and $\bf{n}^{l} \sim\mathcal{CN}(\mathbf{0},\sigma^2\bm{I})$ is an additive white Gaussian noise (AWGN) vector. Without loss of generality, we assume $\frac{1}{N}{\rm tr}\{\bf{A}^{\rm{H}}\bf{A}\}=1$, and define the signal-to-noise ratio (SNR) as ${snr} = \sigma^{-2}$.

Based on $\bf{y}^l$, an iterative receiver is implemented to recover message sequence $\bf{d}$, which consists of an LD and an NLD. The LD corresponds to the linear constraint~\eqref{Eqn:y^t}, and the NLD, consisting of a demodulator and an \emph{a-posteriori probability} (APP) decoder, corresponds to the FEC coding constraint $\bf{x} \in \mathcal{C}$ ($\mathcal{C}$ is the set of codewords). To be specific, as shown in Fig.~\ref{Fig:System model}, based on $\bf{y}^l$ and $\bf{x}^l_t=[x^{l,1}_t,\cdots,x^{l,N}_t]^{\rm{T}}$, the output estimation of LD is $\bf{r}^l_t=[r^{l,1}_t,\cdots,r^{l,N}_t]^{\rm{T}}$. After parallel-to-serial conversion, $\bf{r}_t=[\bf{r}^{1^{\rm{T}}}_t,\cdots,\bf{r}^{L^{\rm{T}}}_t]^{\rm{T}}$ is input to the NLD at $t$-th iteration and then the updated estimation $\bf{x}_{t+1}=[\bf{x}^{1^{\rm{T}}}_{t+1},\cdots,\bf{x}^{L^{\rm{T}}}_{t+1}]^{\rm{T}}$ is fed back to the LD.
The iterative process stops when message sequence $\hat{\bf{d}}$ is recovered successfully or the maximum number of iterations is reached.

The GMIMO system satisfies the following assumptions.
\begin{itemize}
	\item There are huge numbers of transmit and receive antennas, i.e., $N,M \rightarrow \infty$ and channel load $\beta=N/M$ is fixed.
	\item The entries of signal $\bf{x}$ are taken from an arbitrary distribution (e.g., quadrature phase-shift keying (QPSK), quadrature amplitude modulation (QAM), Gaussian, Bernoulli-Gaussian, etc.).
	\item Channel matrix $\bf{A}$ is right-unitarily-invariant, covering various types of channel matrices, e.g., IID random matrices (i.e., Rayleigh fading matrices), certain ill-conditioned and correlated matrices\cite{MaTWC2019}. Let the SVD of $\bf{A}$ be $\bf{A} = \bf{U}\bf{\Lambda}\bf{V}^{\rm{H}}$, where $\bf{U} \in \mathbb{C}^{M\times M}$ and $\bf{V} \in \mathbb{C}^{N\times N}$ are unitary  matrices, and $\bf{\Lambda}\in \mathbb{C}^{M\times N}$ is a rectangular diagonal matrix. $\bf{U}\bf{\Lambda}$ and $\bf{V}$ are independent, and $\bf{V}$ is Haar-distributed (uniformly distributed over all unitary matrices)\cite{RandomWire}. 
    \item Channel matrix $\bf{A}$ is only available to the receiver but unknown to the transmitter.
\end{itemize}

\section{MAMP Receiver and State Evolution}
Since the detection process of \eqref{Eqn:y^t} in each time slot is the same, the time index $l$ is omitted in the rest of this paper for simplicity. Then, the received signal in \eqref{Eqn:y^t} can be rewritten~as:
{\setlength\abovedisplayskip{0.01cm}
\setlength\belowdisplayskip{0.01cm}
\BS\label{Eqn:y}
\begin{align}
\text{Linear constraint}\quad \Gamma:& \;\;\; \bf{y} = \bf{A}\bf{x} + \bf{n},\\
\text{Code constraint}\quad \Phi_{\mathcal{C}}: &\;\;\; \bf{x} \in \mathcal{C} \;\; {\mr{and}}\;\; x_i \sim P_X(x_i), \forall i.
\end{align}
\ES
}

\begin{figure}[!t]
    \centering
    \subfigure[MAMP receiver: $\hat{\gamma}_t$ and $\hat{\phi}_t$ denote the LM-MF detector and demodulator and APP decoder for local constraints $\Gamma$ and $\Phi_{\mathcal{C}}$, respectively.]
    {
        \begin{minipage}[t]{0.9\linewidth}\label{Fig:MAMP receiver} 
            \centering
            \includegraphics[width=0.96\linewidth]{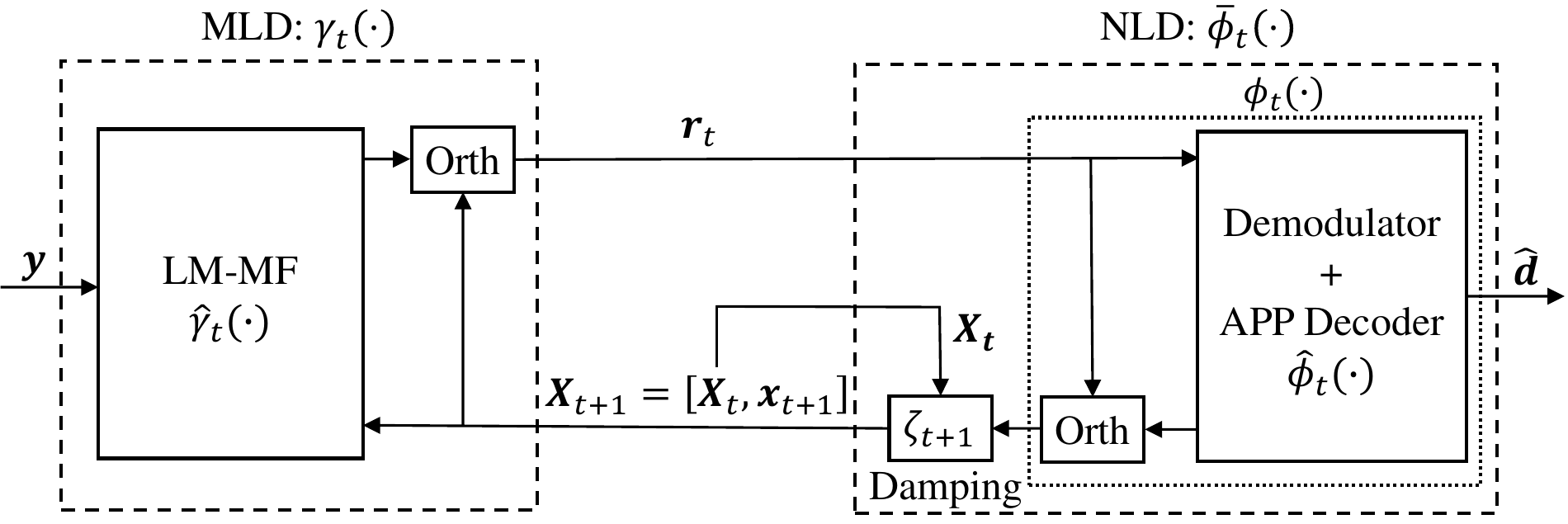}\vspace{0.2cm}
		\end{minipage}
    }\\    
    \subfigure[Transfer functions: $\gamma_{\rm SE}$ and $\bar{\phi}_{\rm SE}$ are the MSE transfer functions of $\gamma_t$ and $\bar{\phi}_t$, respectively.]
	{
		\begin{minipage}[t]{0.9\linewidth}\label{Fig:MAMP SE} 
			\centering
			\includegraphics[width=0.83\linewidth]{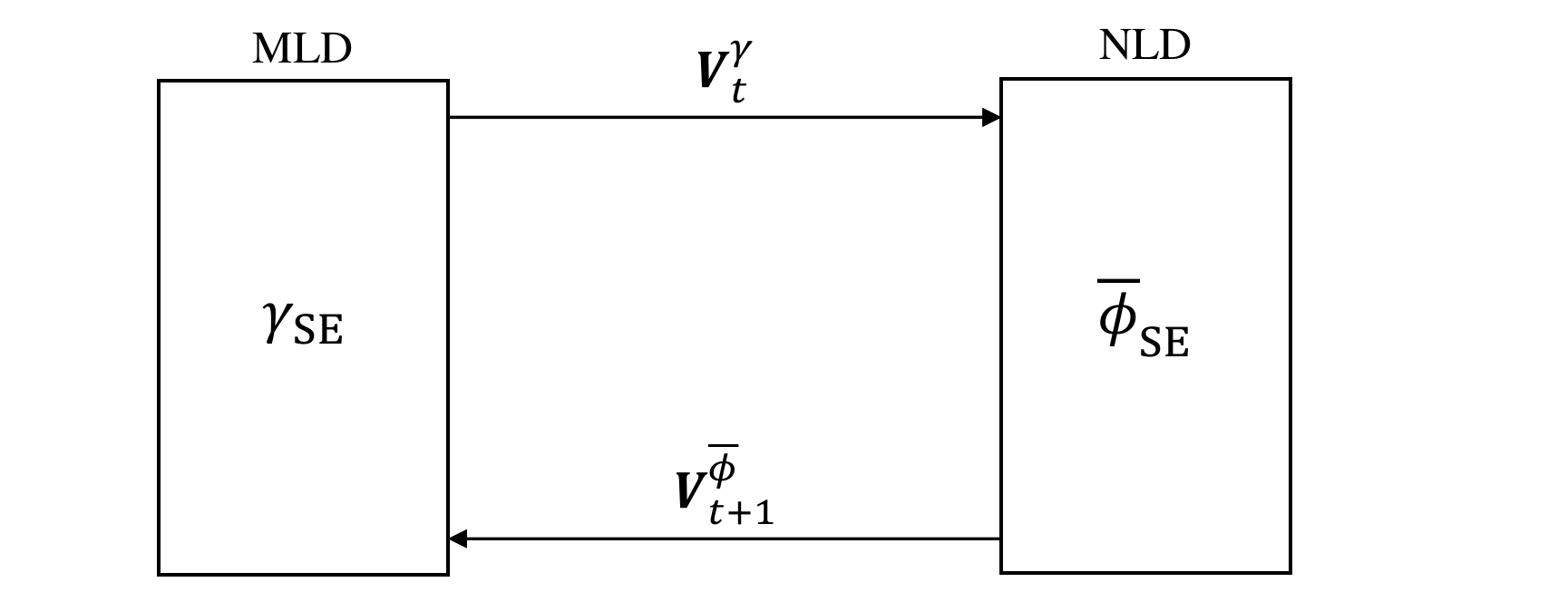}\vspace{0.2cm}
		\end{minipage} 
	}\vspace{-0.2cm}
	\caption{Illustration of the MAMP receiver and its state evolution.}\vspace{-0.3cm}
\end{figure}

\subsection{MAMP Receiver}
Based on \eqref{Eqn:y}, Fig. \ref{Fig:MAMP receiver} shows the MAMP receiver for coded GMIMO, i.e., starting with $t=1$ and $\bf{X}_1=\bf{0}$,
\BS\label{Eqn:BO-MAMP}\begin{align}
	\!\!\mathrm{MLD:}\;\;\;\;\bf{r}_{t} &= \gamma_t(\bf{X}_t) = \tfrac{1}{\varepsilon^{\gamma}_t}(\hat{\gamma}_{t}(\bf{X}_t)-\bf{X}_t\bf{p}_t )  ,\label{Eqn:MAMP-MLE}\\
	\!\!\mathrm{NLD:}\;\bf{x}_{t+1} &= \bar{\phi}_t(\bf{r}_t) = [\bf{X}_t,\phi_t(\bf{r}_t)]\cdot \bf{\zeta}_{t+1}.\label{Eqn:MAMP-NLE}
\end{align}  
\ES

In~\eqref{Eqn:MAMP-MLE}, $\bf{X}_t=[\bf{x}_1,...,\bf{x}_t]$, the local LM-MF $\hat{\gamma}_{t}(\cdot)$ corresponds to the linear constraint $\Gamma$, and the normalized parameters $\{\varepsilon^{\gamma}_t\}$ and orthogonal parameters $\{\bf{p}_t\}$ are designed to ensure the orthogonality for MAMP~\cite{MAMPTIT}. Specifically, $\hat{\gamma}_{t}(\cdot)$ is defined as
\BE\label{Eqn:localLMMF}
\hat{\gamma}_{t}(\bf{X}_t)=\bf{A}^{\rm{H}}\tilde{\gamma}_t(\bf{X}_t), 
\EE
with $\tilde{\gamma}_t(\bf{X}_t)=\theta_t \bf{B} \tilde{\gamma}_{t-1}(\bf{X}_{t-1}) + \xi_t(\bf{y}-\bf{A}\bf{x}_t)$,
where $\tilde{\gamma}_{0}(\bf{X}_{0})=\bf{0}$, $\bf{B}=\lambda^{\dagger}\bf{I}-\bf{A}\bf{A}^{\rm{H}}$ with $\lambda^{\dagger}=(\lambda_{\text{min}}+ \lambda_{\text{max}})/2$, $\lambda_{\text{min}}$ and $\lambda_{\text{max}}$ denote the minimal and maximal eigenvalues of $\bf{A}\bf{A}^{\rm{H}}$, and relaxation parameters $\{\theta_t\}$ and weights $\{\xi_t\}$ can be optimized to improve the convergence speed of the MAMP receiver\cite{MAMPTIT}. Noting that when $\tilde{\gamma}_t(\cdot)$ tends to converge, the output of $\hat{\gamma}_{t}(\bf{X}_t)$ converges to the LMMSE estimation in the LD of OAMP/VAMP\cite{LeiOptOAMP}.

In~\eqref{Eqn:MAMP-NLE}, $\phi_t(\bf{r}_t) \equiv \tfrac{1}{\varepsilon^{\phi}_t}(\hat{\phi}_t(\bf{r}_t)-w_t\bf{r}_t)$, where the local detector $\hat{\phi}_t(\cdot)$ corresponds to the code constraint $\Phi_{\mathcal{C}}$, and parameters $\{\varepsilon^{\gamma}_t\}$ and $\{w_t\}$ are utilized to guarantee the orthogonality for MAMP~\cite{MAMPTIT}. Specifically, $\hat{\phi}_t(\cdot)$ is defined as
\BE\label{Eqn:decoder}
\hat{\phi}_t(\bf{r}_t)\equiv\mr{E}\{\bf{x} |\bf{r}_t,\Phi_{\mathcal{C}}\},
\EE
which corresponds to the demodulation and APP channel decoding \cite[Equation(10)]{MaTWC2019}. It is noted that $\hat{\phi}_t(\cdot)$ is assumed to be Lipschitz-continuous in this paper. Meanwhile, a damping vector $\bf{\zeta}_{t+1} = [\zeta_{t+1,1},...,\zeta_{t+1,t+1}]^\mathrm{T}$ with $\textstyle \sum_{i=1}^{t+1}\zeta_{t+1,i}=1$ is employed to guarantee and improve MAMP convergence while also preserving orthogonality, as demonstrated in \cite{MAMPTIT}.

\vspace{-0.05cm}
\subsection{State Evolution (SE)}
Since the LM-MF is employed in MLD, a covariance-matrix SE is required to evaluate the asymptotic performance.

Define the covariance matrices as:
\BE\label{Eqn:cor-matrix}
    \bf{V}^{\gamma}_{t} \equiv  [v^{\gamma}_{i,j}]_{t\times t} , \;\;\;\;\;\;
    \bf{V}^{\bar{\phi}}_{t} \equiv  [v^{\bar{\phi}}_{i,j}]_{t\times t} ,
\EE
where $v^{\gamma}_{t,t^{\prime}}=(v^{\gamma}_{t^{\prime},t})^{*}\equiv \tfrac{1}{N}\mr{E}\{\bf{g}^{\rm H}_t \bf{g}^{\prime}_t\}$ with $\bf{g}_t=\bf{r}_t-\bf{x}$, $v^{\bar{\phi}}_{t,t^{\prime}}=(v^{\bar{\phi}}_{t^{\prime},t})^{*}\equiv \tfrac{1}{N}\mr{E}\{\bf{f}^{\rm H}_t \bf{f}^{\prime}_t\}$ with $\bf{f}_t=\bf{x}_t-\bf{x}$, and $1\le t^{\prime}\le t$.

Based on the orthogonality and IID Gaussianity property~\cite{MAMPTIT}, the asymptotic MSE performance
of MAMP can be predicted by the MSE functions $\gamma_{\rm SE}(\cdot)$ and $\bar{\phi}_{\rm SE}(\cdot)$, i.e., 
\BS\label{Eqn:SE}
\begin{align}
\mathrm{MLD:}\;\;\;\;\;\;\;\bf{V}^{\gamma}_{t} &= \gamma_{\rm SE}(\bf{V}^{\bar{\phi}}_{t}),\label{Eqn:SE-MLE}\\
\mathrm{NLD:}\;\;\;\;\;\bf{V}^{\bar{\phi}}_{t+1} &= \bar{\phi}_{\rm SE}(\bf{V}^{\gamma}_{t}),\label{Eqn:SE-NLE}
\end{align}
\ES
where $\bf{V}^{\gamma}_{t}$ and $\bf{V}^{\bar{\phi}}_{t}$ are  defined in \eqref{Eqn:cor-matrix}, and $\gamma_{\rm SE}(\cdot)$ and $\bar{\phi}_{\rm SE}(\cdot)$ correspond to constraints \eqref{Eqn:MAMP-MLE} and \eqref{Eqn:MAMP-NLE}, respectively. Moreover, Fig.~\ref{Fig:MAMP SE} gives a graphical illustration of the SE in \eqref{Eqn:SE}.

It is worth noting that the MSE functions in \eqref{Eqn:SE} are multi-dimensional. Therefore, it is very challenging to design FEC codes and analyze the information-theoretic limit for MAMP.

\emph{Note:} It has been proved that $\hat{\gamma}_t(\cdot)$ is Lipschitz-continuous in \cite{MAMPTIT}. The $\hat{\phi}_t(\cdot)$ corresponds to the coding constraint for coded GMIMO. Since the LDPC decoder is proved to be Lipschitz-continuous in \cite[Appendix B]{LC-LDPC}, the SE holds for MAMP receiver with LDPC decoding $\hat{\phi}_t(\cdot)$. Therefore, based on SE, a kind of LDPC code is designed for MAMP receiver in simulation results. Although there is no strict proof for other types of FEC codes, we conjecture that $\hat{\phi}_t(\cdot)$ is also Lipschitz-continuous for the majority of FEC codes (e.g., Turbo code, Polar code, Reed-solomon (RS) code, etc.).

\vspace{-0.1cm}
\section{Information-Theoretic Optimality of MAMP}
In this section, we present the achievable rate analysis and optimal coding principle of MAMP. The information-theoretic optimality proof for MAMP is also presented.

\subsection{Achievable Rate Analysis and Coding Principle}
To circumvent the complex multi-dimensional SE analysis of MAMP, we first provide the fixed-point consistency of MAMP and OAMP/VAMP as follows.

\begin{lemma}[Fixed-Point Consistency]\label{lem:same_fp}
Let the SE fixed point of MAMP in \eqref{Eqn:SE} be $(v_{*}^{\gamma}, v_{*}^{\bar{\phi}})$, where $v_{*}^{\gamma}=\lim\limits_{t \to \infty } v_{t,t}^{\gamma}$ and $v_{*}^{\bar{\phi}}=\lim\limits_{t \to \infty } v_{t,t}^{\bar{\phi}}$. MAMP and OAMP/VAMP have the same SE fixed point $(v_{*}^{\gamma}, v_{*}^{\bar{\phi}})$ for arbitrary fixed Lipschitz-continuous $\hat{\phi}_t(\cdot)$.
\end{lemma}

Based on Lemma~\ref{lem:same_fp}, the multi-dimensional SE of MAMP can converge to the same SE fixed point as the SISO SE of OAMP/VAMP for the same APP decoder $\hat{\phi}_t(\cdot)$. This inspires us to attempt to analyze the achievable rate of MAMP with the aid of the SISO SE of OAMP/VAMP.

Specifically, an equivalent transformation of MAMP is obtained by incorporating all orthogonal and damping operations into the MLD, in which the equivalent MAMP is given by
\BS\label{Eqn:BO-MAMP1}
\begin{align}
\mathrm{MLD}: \quad \;\;\;\; \bf{r}_t = \eta_t(\hat{\bf{x}}_t), \\
\mathrm{NLD}: \quad  \hat{\bf{x}}_{t+1} = \hat{\phi}_t(\bf{r}_t),
\end{align}
\ES
where $\eta_t(\cdot)$ is a multi-dimensional MLD involving $\gamma_t(\cdot)$ in \eqref{Eqn:MAMP-MLE}, damping, and orthogonal operations, and $\hat{\bf{x}}_t$ denotes the output \emph{a posteriori} estimation of APP decoder $\hat{\phi}_t(\cdot)$. Note that this equivalent transformation does not change the SE fixed point (i.e., convergence performance) of MAMP. As a result, the equivalent MAMP is also referred to as MAMP for simplicity. In contrast to the multi-dimensional transfer function in \eqref{Eqn:SE-NLE}, the transfer function of $\hat{\phi}_t(\cdot)$ in NLD is SISO. However, since the memory is required in $\eta_t(\cdot)$, the transfer function of $\eta_t(\cdot)$ remains intricately multi-dimensional. This continues to impede the theoretical analysis of MAMP.

To overcome the above issue, a SISO variational transfer function $\eta_{\rm{SE}}$ of MLD $\eta_t(\cdot)$ is derived with the aid of Lemma~\ref{lem:same_fp} and the SE of OAMP/VAMP. Therefore, a SISO variational SE (VSE) of MAMP is presented in the following lemma, which is adopted to simplify the achievable rate analysis and optimal code design for MAMP.

\begin{lemma}[VSE of MAMP]\label{lem:VSE_MAMP}
Let $\rho_t^{\gamma} = 1/ v_{t,t}^{\gamma}$ denote the input signal-to-interference-plus-noise ratio (SINR) of the NLD, the VSE of MAMP can be written as 
\BS\label{Eqn:VSE_MAMP}
\begin{align}
   \!\!\! \mathrm{MLD}:\;\; \;\; \rho^{\gamma}_{t} &= \eta_{\rm{SE}}(v^{\hat{\phi}}_t) = (v^{\hat{\phi}}_t)^{-1}-[\hat{\gamma}_{\mr{SE}}^{-1}(v^{\hat{\phi}}_t)]^{-1} ,\label{Eqn:VSE-MLD}\\
   \!\!\! \mathrm{NLD}:\; v^{\hat{\phi}}_{t+1} &= \hat{\phi}^{\mathcal{C}}_{\rm{SE}}(\rho^{\gamma}_{t}) = \mr{mmse} \{\bf{x}|\sqrt{\rho^{\gamma}_{t}}\bf{x}+\bf{z}, \Phi_{\mathcal{C}}\},\label{Eqn:VSE-NLD}
\end{align}
\ES
where $\hat{\gamma}_{\mr{SE}}(v)= \tfrac{1}{N}{\rm{tr}}\{[snr\bf{A}^{\rm{H}}\bf{A}+v^{-1}\bf{I}]^{-1}\}$ denotes the MSE function of LMMSE detector, $\hat{\gamma}_{\mr{SE}}^{-1}(\cdot)$ the inverse of $\hat{\gamma}_{\mr{SE}}(\cdot)$, and $\bf{z}\sim \mathcal{CN}(\bf{0}, \bf{I})$ an AWGN vector independent of $\bf{x}$.
\end{lemma}

Note that the VSE transfer functions in~\eqref{Eqn:VSE_MAMP} are not equivalent to the SE transfer functions in~\eqref{Eqn:SE}. Although VSE cannot be utilized to characterize the MSE performance of MAMP in each iteration, it can be employed to accurately analyze the achievable rate and coding principle.

Due to the coding gain, the decoding transfer function $\hat{\phi}_{\mr{SE}}^{\mathcal{C}}(\cdot)$ is upper bounded by the demodulation transfer function $\hat{\phi}_{\mr{SE}}^{\mathcal{S}}(\cdot)$, i.e.,
\BE\label{Eqn:CodeMAMP1}
\hat{\phi}^{\mathcal{C}}_{\rm{SE}}(\rho^{\gamma}_{t}) < \hat{\phi}^{\mathcal{S}}_{\rm{SE}}(\rho^{\gamma}_{t}),  \quad \mr{for}\;\; 0\leq \rho^{\gamma}_{t} \leq \rho_{\rm{max}},	\EE
where $\hat{\phi}^{\mathcal{S}}_{\rm{SE}}(\rho^{\gamma}_{t}) = \mr{mmse} \{\bf{x}|\sqrt{\rho^{\gamma}_{t}}\bf{x}+\bf{z}, \Phi_{\mathcal{S}}\}$ and $\rho_{\rm{max}} = snr$

\begin{figure}[!tbp]
	\centering
	\includegraphics[width=0.75\linewidth]{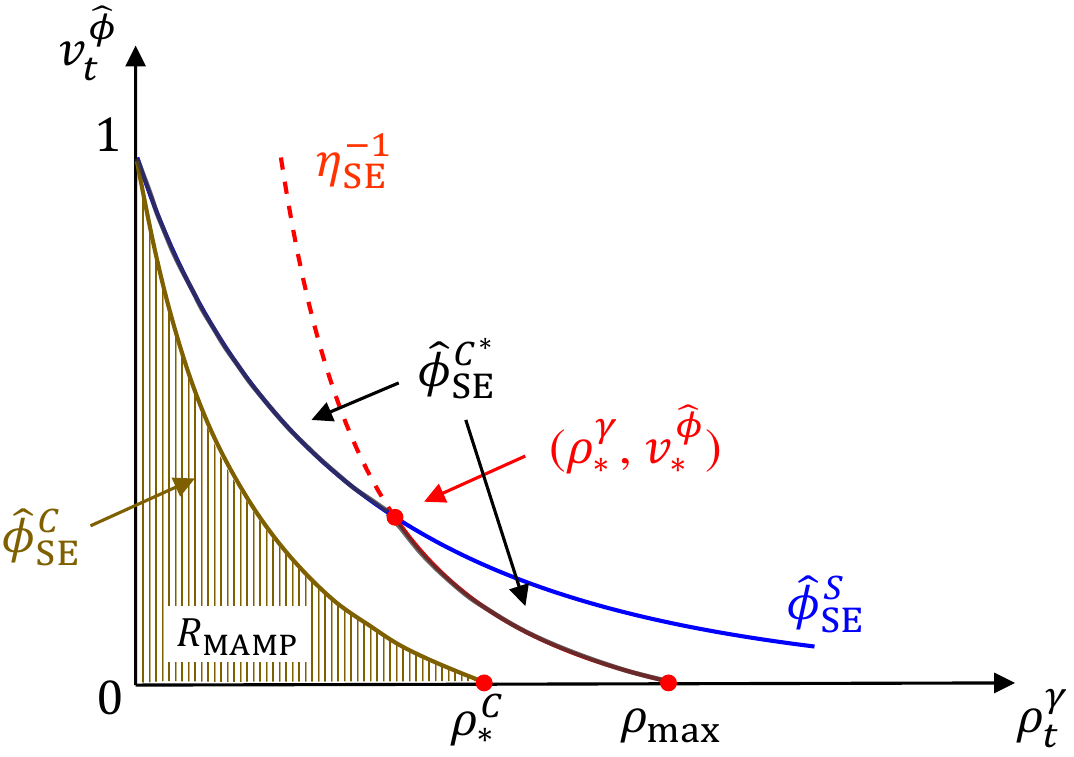}\vspace{-0.3cm}
	\caption{Graphical illustration of VSE for MAMP, where $\eta^{-1}_{\rm{SE}}(\cdot)$ is the inverse function of $\eta_{\rm{SE}}(\cdot)$. $\hat{\phi}^{\mathcal{S}}_{\rm{SE}}(\cdot)$ and $\hat{\phi}^{\mathcal{C}}_{\rm{SE}}(\cdot)$ with $\hat{\phi}_{\mr{SE}}^{\mathcal{C}}(\rho_{*}^{\mathcal{C}})=0$ denote the MMSE functions of constellation and code constraint in NLD, respectively. $(\rho_*^{\gamma}, v_*^{\hat{\phi}})$ denotes the VSE fixed point between $\eta^{-1}_{\rm{SE}}(\cdot)$ and $\hat{\phi}^{\mathcal{S}}_{\rm{SE}}(\cdot)$. Moreover, $\hat{\phi}^{\mathcal{C}^*}_{\rm{SE}}(\cdot)$ is the optimal coding function of MAMP.}\label{Fig:SE curve} \vspace{-0.4cm}
\end{figure}

As shown in Fig.~\ref{Fig:SE curve}, assume that there is a unique fixed point $(\rho_*^{\gamma}, v_*^{\hat{\phi}})$ between $\eta_{\rm{SE}}^{-1}(\cdot)$ and $\hat{\phi}^{\mathcal{S}}_{\rm{SE}}(\cdot)$. Since $v_*^{\hat{\phi}}>0$, the converge performance of MAMP is not error-free. 

Therefore, to achieve the error-free performance, a kind of proper FEC code should be well-designed to guarantee an available decoding tunnel between $\eta_{\rm{SE}}^{-1}(\cdot)$ and $\hat{\phi}^{\mathcal{C}}_{\rm{SE}}(\cdot)$, i.e., 
\BE\label{Eqn:CodeMAMP2}
    \hat{\phi}^{\mathcal{C}}_{\rm{SE}}(\rho^{\gamma}_{t}) < \eta_{\rm{SE}}^{-1}(\rho^{\gamma}_{t}), \quad \mr{for}\;\; 0 \leq \rho^{\gamma}_{t} \leq \rho_{\rm{max}}.
\EE

Therefore, based on \eqref{Eqn:CodeMAMP1} and \eqref{Eqn:CodeMAMP2}, we obtain the error-free condition of MAMP in the following lemma.
\begin{lemma}[Error-Free Decoding]\label{lem:err_free}
MAMP can achieve error-free decoding if and only if 
\BE\label{Eqn:CodeMAMP3}
\hat{\phi}^{\mathcal{C}}_{\rm{SE}}(\rho^{\gamma}_{t}) <  \min\{\hat{\phi}^{\mathcal{S}}_{\rm{SE}}(\rho^{\gamma}_{t}), \eta_{\rm{SE}}^{-1}(\rho^{\gamma}_{t})\},  \quad \mr{for}\;\; 0\leq \rho^{\gamma}_{t} \leq \rho_{\mr{max}}.
\EE
\end{lemma}

Then, based on Lemma~\ref{lem:err_free} and I-MMSE lemma\cite{GuoTIT2005}, we give the achievable rate of MAMP as follows.

\begin{lemma}[Achievable Rate of MAMP]\label{lem:Rate}
The achievable rate of MAMP with fixed $\hat{\phi}_{\mr{SE}}^{\mathcal{C}}(\cdot)$ is
\BE \label{Eqn:rate_MAMP}
\begin{aligned}
 &R_{\text{MAMP}} = \int_{0}^{\rho_{*}^{\mathcal{C}}}
 \hat{\phi}_{\mr{SE}}^{\mathcal{C}}(\rho_t^{\gamma}) d \rho_t^{\gamma},   \\
 &\begin{array}{l@{\quad}l}
 {\rm s.t.} &  \hat{\phi}_{\mr{SE}}^{\mathcal{C}}(\rho_t^{\gamma})< \hat{\phi}_{\mr{SE}}^{\mathcal{C}^*}(\rho_t^{\gamma}), \;\;  \mr{for}\;\;\;0\le \rho_t^{\gamma} \le \rho_{\mr{max}},
 \end{array}
\end{aligned}
\EE    
where $\hat{\phi}_{\mr{SE}}^{\mathcal{C}^*}(\rho_t^{\gamma})=\mr{min}\{\hat{\phi}^{\mathcal{S}}_{\rm{SE}}(\rho_t^{\gamma}), \eta_{\rm{SE}}^{-1}(\rho_t^{\gamma})\}$ and $\rho_{*}^{\mathcal{C}}=\hat{\phi}_{\mr{SE}}^{\mathcal{C}^{-1}}(0)$.
\end{lemma}

Therefore, based on Lemma~\ref{lem:Rate}, the optimal code design principle of MAMP can be obtained in the following lemma.

\begin{lemma}[Optimal Code Design]\label{lem:optimal code design}
    The optimal coding principle of MAMP is
    \BE\label{Eqn: Optimal coded desdign}
    \hat{\phi}_{\mr{SE}}^{\mathcal{C}}(\rho_t^{\gamma}) \to \hat{\phi}_{\mr{SE}}^{\mathcal{C}^*}(\rho_t^{\gamma}), \quad \mr{for}\;\;\; 0\le \rho_t^{\gamma} \le \rho_{\mr{max}}, 
    \EE
    enabling MAMP to achieve error-free performance as well as the maximum achievable rate.
\end{lemma}

\begin{figure}\vspace{-0.3cm}
    \centering
    \includegraphics[width=0.92\linewidth]{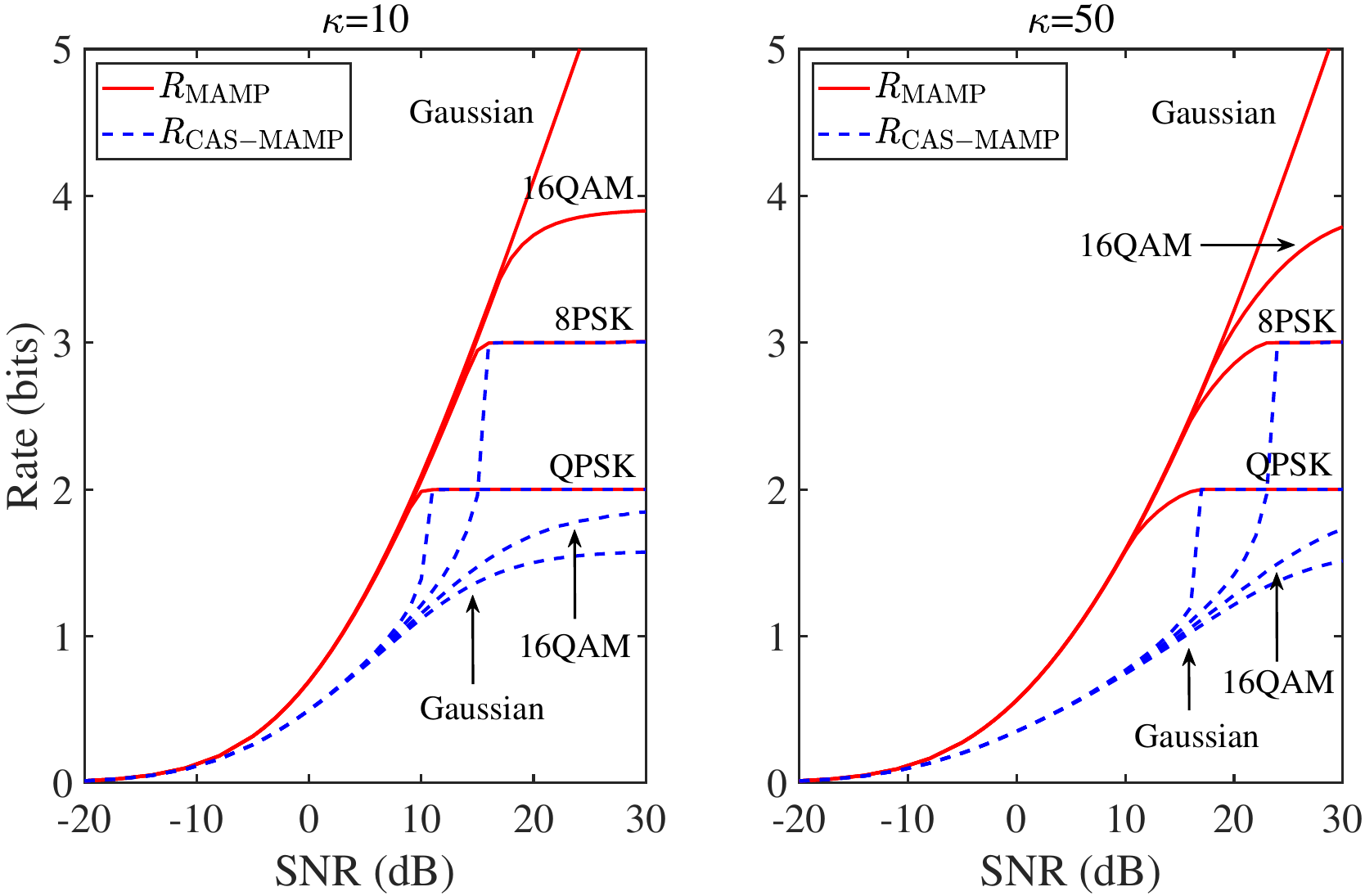}\vspace{-0.25cm}
	
    \caption{Maximum achievable rate comparison between MAMP and CAS-MAMP for GMIMO with $\bf{A}_{\rm ill}$, where $N=500$, $\beta=1.5$, and $\kappa=\{10,50\}$.}\label{Fig:AR} \vspace{-0.3cm}
\end{figure}

\vspace{-0.1cm}
\subsection{Information-Theoretic Optimality Proof of MAMP}\label{sec:IT-MAMP}

Based on Lemma~\ref{lem:optimal code design}, the maximum achievable rate of MAMP is obtained directly in the following theorem.

\begin{theorem}[Maximum Achievable Rate]\label{The:maxrate}
The maximum achievable rate of MAMP is 
\BE\label{Eqn:rateMAMP}
  	R_{\rm{MAMP}}^{\rm{max}} \to  \int_{0}^{\rho_{\rm{max}}}  \hat{\phi}^{\mathcal{C}^*}_{\rm{SE}}(\rho_t^{\gamma}) d\rho_t^{\gamma},
\EE
where $\hat{\phi}^{\mathcal{C}^*}_{\rm{SE}}(\rho_t^{\gamma})={\rm{min}}\{\hat{\phi}^{\mathcal{S}}_{\rm{SE}}(\rho_t^{\gamma}), \eta_{\rm{SE}}^{-1}(\rho_t^{\gamma})\}$.
\end{theorem}

Due to the constrained-capacity optimality of OAMP/VAMP\cite{LeiOptOAMP,YuhaoTcom2022}, based on Lemma~\ref{lem:VSE_MAMP}, Lemma~\ref{lem:optimal code design}, and Theorem~\ref{The:maxrate}, the information-theoretic (i.e., constrained-capacity) optimality of MAMP is verified in the following theorem.
\begin{theorem}[Constrained-Capacity Optimality]\label{The:Constrained-Capacity Optimality}
MAMP can achieve the same maximum achievable rate as OAMP/VAMP, which indicates MAMP is constrained-capacity optimal in GMIMO, i.e.,
\BS\label{Eqn:capcacity}
\begin{align}
	   &R_{\rm{MAMP}}^{\rm{max}} = R_{\rm{OAMP/VAMP}}^{\rm{max}}\to \int_{0}^{\rho_{\rm{max}}}  \hat{\phi}^{\mathcal{C}^*}_{\rm{SE}}(\rho) d\rho,   \\
         {\rm s.t.} &  \quad \hat{\phi}^{\mathcal{C}}_{\rm{SE}}(\rho) \to \hat{\phi}^{\mathcal{C}^*}_{\rm{SE}}(\rho) = \rm{min}\{ \hat{\phi}^{\mathcal{S}}_{\rm{SE}}(\rho), \eta^{-1}_{\rm{SE}}(\rho)\},
\end{align}
\ES
where $R_{\rm{MAMP}}^{\rm{max}}$ is equal to the constrained capacity of GMIMO given in \cite{LeiOptOAMP,YuhaoTcom2022}.
\end{theorem}

To demonstrate the advantages of MAMP, we also present the maximum achievable rates of the conventional CAS-MAMP receiver~\cite{CDMA2005}, in which the MLD and NLD are implemented sequentially without iteration over each other. Based on I-MMSE lemma\cite{GuoTIT2005}, Fig.~\ref{Fig:SE curve} shows that the maximum achievable rate of CAS-MAMP is $R_{\rm{CAS-MAMP}}^{\text{max}} = \int_{0}^{\rho^{\gamma}_{*}}  \hat{\phi}^{\mathcal{C}^*}_{\rm{SE}}(\rho_t^{\gamma}) d\rho_t^{\gamma}$ for given snr.
Therefore, compared with $R^{\text{max}}_{\rm MAMP}$, the rate loss of CAS-MAMP is $ R_{\rm{loss}}= \int_{\rho^{\gamma}_{*}}^{\rho^{\text{max}}}  \hat{\phi}^{\mathcal{C}^*}_{\rm{SE}}(\rho_t^{\gamma}) d\rho_t^{\gamma}$.

Taking ill-conditioned channel matrices $\bf{A}_{\rm ill}$ as examples, Fig.~\ref{Fig:AR} shows the maximum achievable rate comparison between MAMP and CAS-MAMP.
Let the SVD of $\bf{A}_{\rm ill}$ be $\bf{A}_{\rm ill} = \bf{U}\bf{\Lambda}\bf{V}^{\rm{H}}$. $\bf{U}$ and $\bf{V}$ are generated by unitary matrices of SVD decomposition of an IID Gaussian matrix. We set the eigenvalues $\{e_i\}$ in $\bf{\Lambda}$ as\cite{Vila2015ICASSP}: $e_i/e_{i+1} = \kappa^{1/\mathcal{T}}, i=1,...,\mathcal{T}-1$, and $\sum\nolimits_{i = 1}^{\mathcal{T}} {e_i^2 = N}$, where $\kappa \ge 1$ denotes the condition number of $\bf{A}_{\rm ill}$ and $\mathcal{T}\!=\!\rm{min}\{M,N\}$. As shown in Fig.~\ref{Fig:AR}, the maximum achievable rates of MAMP are higher than those of CAS-MAMP, and increase with the modulation order and SNR. Moreover, different from MAMP, the achievable rate of CAS-MAMP with high-order modulation (e.g., Gaussian signaling and 16QAM) is lower than that of low-order modulation (e.g., QPSK). This similar phenomena has been discussed in \cite{LeiTIT2021}.

\begin{figure}\vspace{-0.3cm}
    \centering
    \includegraphics[width=0.93
    \linewidth]{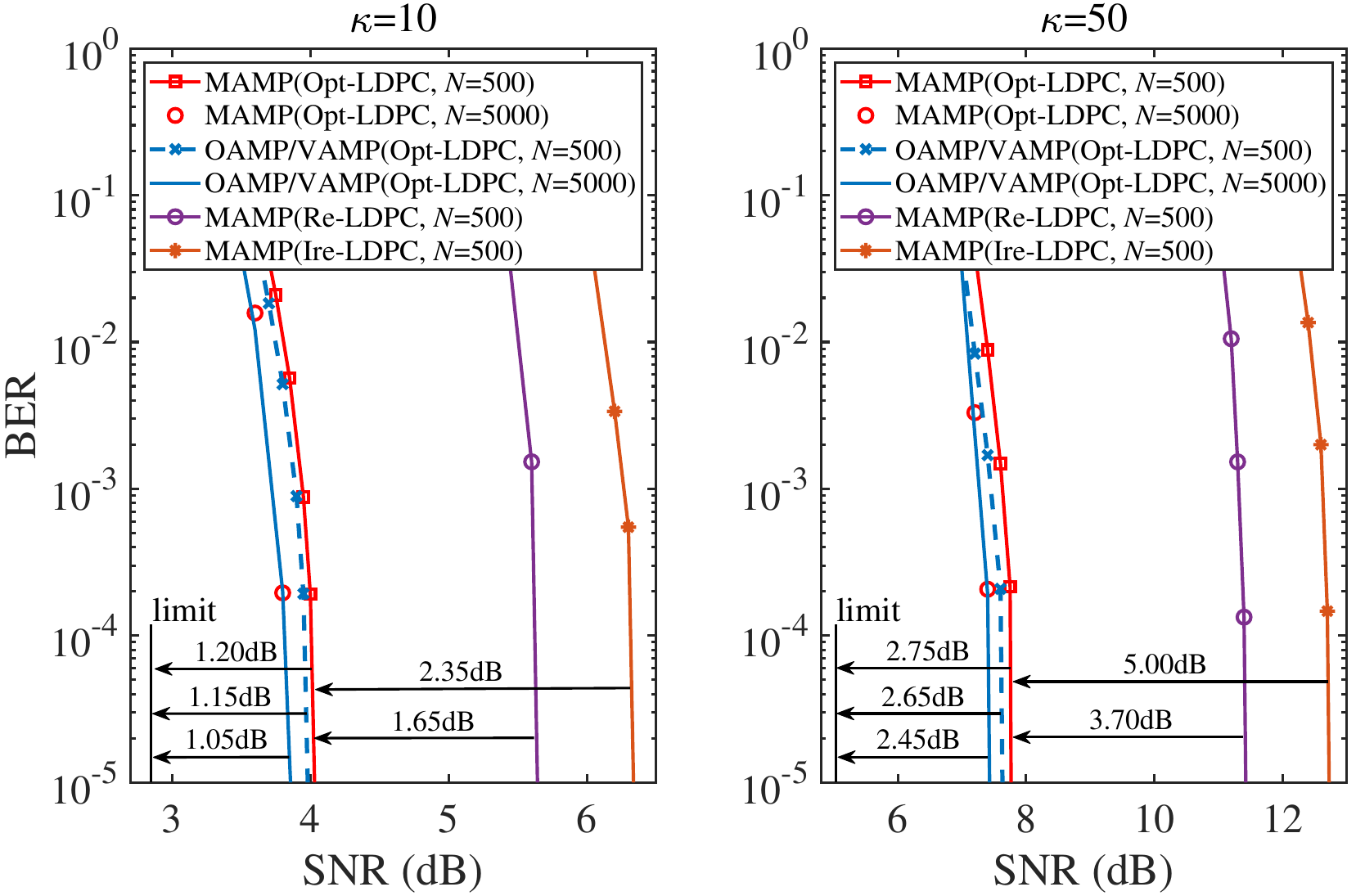}\vspace{-0.3cm}
	
    \caption{BER performance of MAMP and OAMP/VAMP with optimized LDPC codes and MAMP with P2P LDPC codes for $\bf{A}_{\rm ill}$, where $\beta=1.5$, $\kappa=\{10,50\}$, $N=\{500,5000\}$ and code length = $\{1\times10^5,2\times10^5\}$. ``Opt-LDPC'' is the optimized LDPC code, ``Re-LDPC'' the P2P-regular (3,6) LDPC code with code rate $=0.5$~\cite{ryan2009channel}, ``Ire-LDPC'' the well-designed P2P-irregular code with code rate $=0.5$~\cite{Richardson2001}, and ``limit'' the associated constrained capacity.}\label{Fig:BER} \vspace{-0.3cm}
\end{figure}

\section{Numerical Results}\label{sec:results}
In this section, we present the bit error rate (BER) performances and running time complexity of MAMP with optimized LDPC codes in GMIMO. Meanwhile, BER performance comparisons with existing schemes are provided.

We assume channel matrix is ill-conditioned. A kind of irregular LDPC code with code rate $\approx 0.5$ is optimized based on Lemma~\ref{lem:optimal code design} and QPSK modulation is employed.

As shown in Fig.~\ref{Fig:BER}, the BER comparisons between MAMP and OAMP/VAMP with optimized LDPC codes are presented. When $N=500$, the gaps between BER curves at $10^{-5}$ of MAMP and OAMP/VAMP are within $0.1$ dB, since the optimality analysis for MAMP is based on infinite length assumption. Meanwhile, we provide the BER comparisons between MAMP and OAMP/VAMP for large-scale systems, i.e., $N=5000$. It is worth noting that MAMP can achieve the same BER performances as OAMP/VAMP in large-scale systems, but with much lower complexity than OAMP/VAMP.

To validate the advantages of the optimized LDPC codes, the BER performances of MAMP with P2P regular and well-designed irregular LDPC codes are also present. As shown in Fig.~\ref{Fig:BER}, the gaps between BER curves of the optimized LDPC codes and the associated constrained capacities are about $1.0\sim2.7$~dB, which verifies the capacity-approaching performances of the optimized LDPC codes. Moreover, MAMP with the optimized codes have $1.6\sim 5.0$ dB gains over the MAMP with P2P LDPC codes. This indicates that the Bayes-optimal MAMP with well-designed P2P LDPC codes are not optimal anymore with significant performance losses in GMIMO.

\begin{figure}[!tbp]
	\centering
	\includegraphics[width=0.9\linewidth]{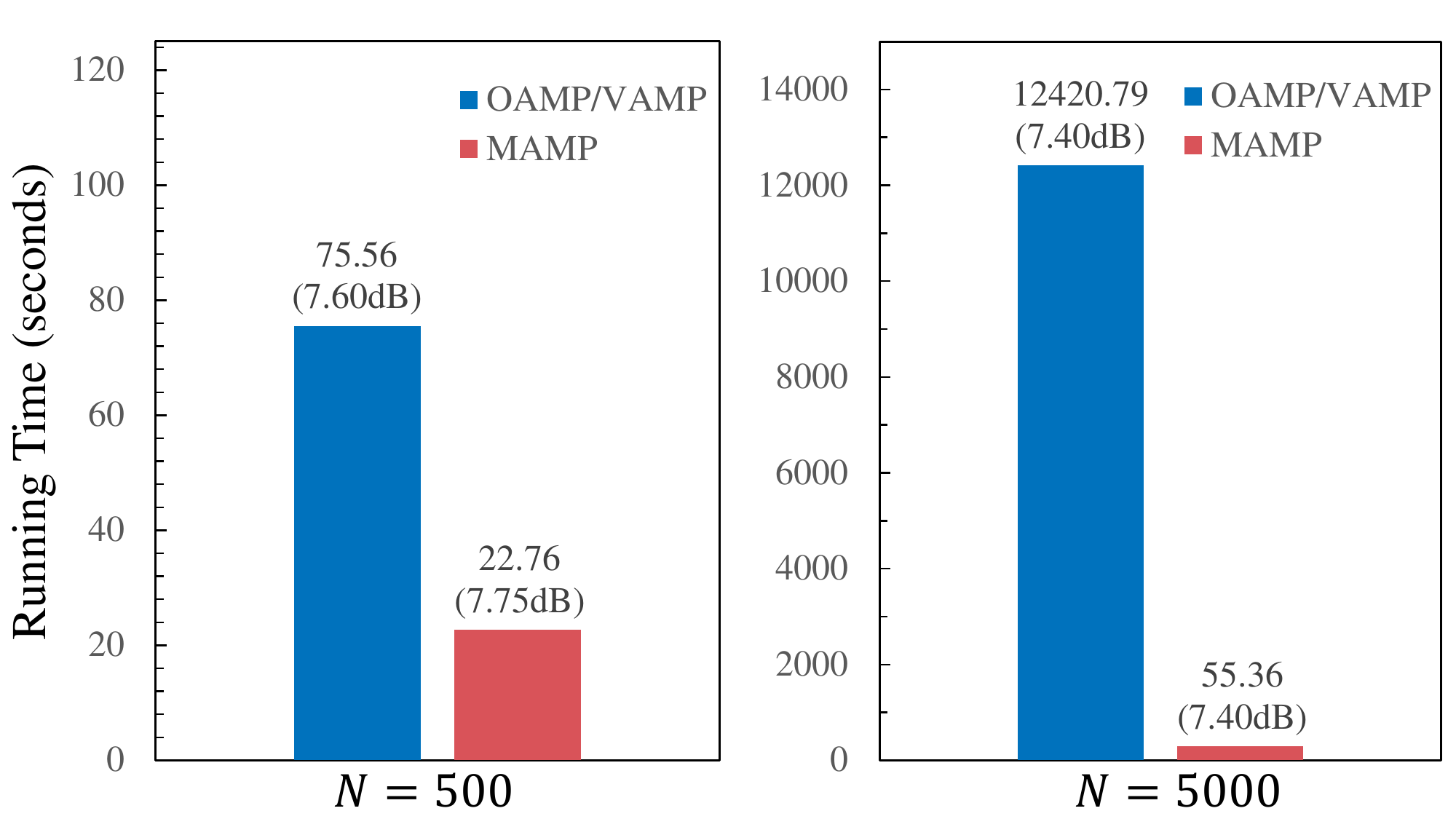}\vspace{-0.3cm}
	
	\caption{Running time comparison between MAMP and OAMP/VAMP, where target ${\text{BER}}=2\times 10^{-4}$, $\kappa=50$, $\beta=1.5$, and $N=\{500,5000\}$ .}\label{Fig:running time} \vspace{-0.4cm}
\end{figure}

To intuitively highlight the low-complexity advantage of MAMP, the running time comparison of MAMP and OAMP/VAMP is shown in Fig.~\ref{Fig:running time}. The running time is obtained by Matlab 2021a on a PC with an Intel Core i7-11700F CPU and 16 GB of RAM. 
For $N=500$, Fig.~\ref{Fig:running time} shows that the running time of MAMP is just $30\%$ of that of OAMP/VAMP. When $N$ increases to 5000, MAMP can achieve the same performance as OAMP/VAMP with only $4\text{\textperthousand}$ of the time consumption.
The reason is that the complexity of MAMP and OAMP/VAMP is determined by MLD with complexity $\mathcal{O}(MN\tau+N\tau^2+\tau^3)$ and LD with complexity $\mathcal{O}\left((M^2N+M^3)\tau\right)$, respectively\cite{MAMPTIT}, where $\tau$ is the number of iterations. The complexity of NLD is identical for MAMP and OAMP/VAMP due to the same demodulation and LDPC decoder employed in NLD.
Therefore, compared with OAMP/VAMP, MAMP can achieve the information-theoretic limit of GMIMO with significantly lower implementation complexity, making it a very promising candidate for large-scale systems.

\vspace{-0.1cm}
\section{Conclusion}
This paper studies the achievable rate analysis and coding principle of the low-complexity MAMP in GMIMO, demonstrating its information-theoretic optimality. To overcome the difficulty in multi-dimensional SE analysis of MAMP, a simplified SISO VSE is derived for MAMP, based on which its achievable rate is calculated and optimal coding principle is established to maximize the achievable rate. 
Moreover, the information-theoretic optimality of MAMP is proved.
Numerical results show that the BER performances of MAMP with optimized LDPC codes approach the constrained capacities and MAMP achieve the same performance with $4\text{\textperthousand}$ of running time compared to OAMP/VAMP for large-scale systems.

\footnotesize
\bibliographystyle{IEEEtran}
\bibliography{manuscript}

\end{document}